# Magnetization reversal more rapidly by using an ultrashort square-wave laser pulse

Xiaoqiang Zhang,[1,2] Yong Xu,[1,2] and Weisheng Zhao[1,2, a)]

[1] Hefei Innovation Research Institute, School of Microelectronics, Beihang University, Hefei 230013, China

[2] Fert Beijing Institute, BDBC, Beihang University, Beijing 100191, China

[a)] Author to whom correspondence should be addressed. Electronic mail: weisheng.zhao@buaa.edu.cn

**Abstract**: With the feature of low-power magnetization manipulation at an ultrashort time scale, all optical switching (AOS) has been propelled to the forefront in investigations. To further speed up the magnetization reversal by manipulating ultrashort optical pulses, in this paper, one single square-wave laser pulse (SWLP) vie the combination of heating and Inverse Faraday Effect (IFE) is explored to excite the reversal of magnetization in a Co/Pt system. Simulation results show that the switching time of magnetization is 3 times faster than the using of a traditional Gaussian wave laser pulse (GWLP) under the same laser energy and pulse duration, and the threshold of AOS for the ferromagnet is 0.67 mJ/cm$^2$. We furthermore demonstrate that the "heat accumulating effect" of laser-pulse is an important factor that influences the switching time, and a SWLP has a larger effect of heat accumulating than a GWLP.

Ultrafast magnetization dynamics is drawing great interest since the observation of a *ps*-time scale demagnetization dynamics in *Ni* films after a *fs*-time scale pulsed-laser excitation.[1] All optical switching (AOS), which have the feature of low-power magnetization manipulation at an ultrashort time scale, has been propelled to the forefront in investigations.[2-4] This novel phenomenon was first observed in a perpendicularly magnetized ferrimagnetic GdFeCo alloy,[5] and later a wider range of materials were discovered.[6,7] One of the most promising applications of AOS is the manipulation and storage of magnetic information, by using laser pulses to switch the magnetization rapidly without any external magnetic field.[8] To further speed up magnetization reversal by manipulating ultrashort optical pulses, triggering magnetization switching on shorter time scales, is a promising strategy.[9] However, conventional Gaussian wave laser pulses (GWLP), as the only one candidate, are applied to excite AOS in the past two decades.[10,11] To switch the magnetization more rapidly, in this manuscript, one single square-wave laser pulse (SWLP) is first employed to excite AOS in a Co/Pt system. Compare with traditional GWLP, where each pulse has a Gaussian temporal profile, a SWLP has a flattop,[12] and the switch time of magnetization is 3 times faster, where is more suitable for the applications of magnetic data storage.

SWLPs, which can be used for potential applications including laser micromachining, optical sensing, and optical square-wave clocks, have recommend themselves to be a topic of great interest in optics,[13] and many new technologies have been proposed to generate SWLPs, e.g. passively mode-locked fiber lasers [14]. With these technologies proliferating, manipulating the reversal of magnetization of ferromagnets by using of a SWLP is possible. Previous studies have confirmed that two main mechanisms, which are inverse Faraday effect (IFE) and thermal effects, are responsible for AOS.[15] And several theories have successfully uncovered the underlying mechanism of AOS, e.g. momentum resolved Boltzmann scattering,[16] atomistic Landau-Lifshitz Gilbert (LLG),[17] and Landau-Lifshitz-Bloch (LLB).[18] Recently, a quite simple model, which is called microscopic three-temperature model (M3TM), is successfully accounted for the AOS of a Co/Pt system.[15] In this model, free electrons (*e*), phonons according to the Einstein or Debye model (*p*), and spin excitation is represented by a mean-field Weiss model (*s*). In this Rapid Communication, M3TM is re-employed to describe AOS exciting by a SWLP in a ferromagnet, and the IFE, which is induced by the helicity of the SWLP, acts as one of the driving mechanisms.

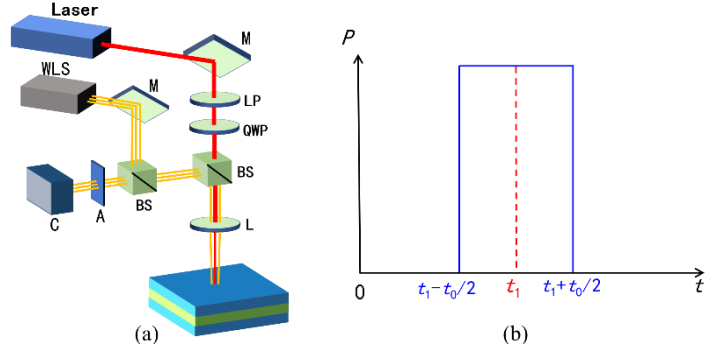

Fig. 1 (a) Sketch of the AOS setup with the a SWLP. Laser, femtosecond square-wave laser; LP, linear polarizer; QWP, quarter wave plate; L, lens; WLS, white light source; M, mirror; A, analyzer, C: CCD camera, BS: beam splitter. (b) The temporal profile of the square-wave laser pulse centered $t = t_1$, and the pulse duration is $t_0$.

Figure 1 (a) shows the sketch of the AOS setup with a SWLP, and a femtosecond square-wave laser pulse with width $t_0 = 35$ *fs* is generated from the femtosecond laser. The polarization of generated SWLP can be controlled by the linear polarizer (LP) and the quarter wave plate (QWP) easily, and "Left" or "right" circular polarization of the pulse can be switched by rotating the QWP by ±45° with respect to the plane of LP. In the sketch, the SWLP is introduced to heat up the electron system of the sample, and then a rapid increase of the electron temperature ($T_e$) and phonon temperature ($T_p$). After a few picoseconds, the heat will dissipate from the electron system into the substrate. The thermodynamic process of electron and lattice can be gotten from the following two differential equations:[11]

$$C_e \frac{dT_e}{dt} = g_{ep}(T_p - T_e) + P(r,t) - \kappa(T_e - T_{amb}) \tag{1.1}$$

$$C_p \frac{dT_p}{dt} = g_{ep}(T_e - T_p), \tag{1.2}$$

where $C_e$ and $C_p$ are the electron and phonon heat capacities, respectively, $g_{ep}$ is the electron-lattice coupling constant, $\kappa$ is the heat diffusion constant, $T_{amb}$ is the ambient temperature, and $P(r, t)$, which is determined by the square-wave laser pulse and the amount of laser energy absorbed by the sample, is the heat source, and $P(r, t)$ has a Gaussian space profile, [11] which coincides with the profile of the laser pulse. In Eq. 1.1 and Eq. 1.2, $C_e$ can be assumed to have a linear approximation of $C_e = \gamma T_e$, where $\gamma$ is a materials-dependent parameter, and $C_p$ is independent of the lattice temperature. In this manuscript, the temporal profile of the SWLP is investigated only, and the Gaussian distribution in space can be ignored when we consider one fixed point of the ferromagnet only. Then, the heat source $P(r, t)$ in Eq. 1.1 can be rewritten as $P(t)$.

As shown in Fig. 1 (b), the fluence of SWLP has a square-wave time-profile, and the heat source can be described as $P_{SWLP}(t) = I_0 \cdot F \cdot rect(t)$, where $I_0$ is assumed to be the amount of laser energy absorbed by the sample [19,20], $F$ is the total fluence of the square-wave pulse, and $rect(t)$ is a square wave. The temporal profile of $rect(t)$ has the following form:

$$rect(t) = \begin{cases} 1 & t_1 - \frac{t_0}{2} < t < t_1 + \frac{t_0}{2}, \\ 0 & else \end{cases} \tag{2}$$

where $t_0$ is the pulse duration and $t_1$ is the center of the pulse duration.

Based on the M3TM, the magnetization dynamics of the spin can be completely specified as follow:[15]

$$\frac{dM}{dt} = \frac{RT_p}{T_c}(M+H_{eff})\left[1-M\cdot\coth\left(\frac{T_c}{T_e}[M+H_{eff}]\right)\right], \quad (3)$$

where $M$ is the magnetization normalized to the saturation value, $H_{eff}$ is the effective magnetic field induced by IFE, and $T_c$ is the Curie temperature. $R$ is the demagnetization rate and $R=\frac{8a_{sf}T_c^2 g_{ep}}{k_B T_D^2 D_S}$, with $a_{sf}$ the spin-flip probability, $k_B$ the Boltzmann constant, $T_D$ the Debye temperature, and $D_s$ the atomic magnetic moment divided by Bohr magneton $\mu_B$. To the best of our knowledge, in experiment, the effective magnetic field induced by IFE is still very difficult to characterize, and the theory accounted for IFE is under development as well. [21-23] Therefore, the strength and the duration of the $H_{eff}$ are estimated basing on existing theories, and given by

$$H_{eff}(t) = \sigma\frac{2\beta F}{ct_0}\cdot f(t)\cdot\vec{k}, \quad (4)$$

where $\sigma$ is the polarization of the SWLP and is equal to $\pm 1$ and 0 for a right-hand or left- hand circularly polarized and linearly polarized light pulse, respectively, $\beta$ is the magneto-optical susceptibility, $c$ is the speed of light, $\vec{k}$ is the unit vector along the wave vector of the SWLP, and $f(t)$ is the temporal profile of induced $H_{eff}$. Similar to Ref. [22], the lifetime of $H_{eff}$ lasts somewhat longer than the laser pulse, and the temporal profile of $f(t)$ can be introduced as

$$f(t) = \begin{cases} 0 & 0 < t < \left(t_1 - \frac{t_0}{2}\right) \\ 1 & \left(t_1 - \frac{t_0}{2}\right) < t < \left(t_1 + \frac{t_0}{2}\right) \\ \exp\left[-\left(\frac{t-(t_1+t_0/2)}{t_0+t_{decay}}\right)^2\right] & t_1 + \frac{t_0}{2} < t < \left(t_1 + \frac{t_0}{2} + t_{decay}\right) \end{cases}, \quad (5)$$

where $t_{decay}$ is defined as the decay time of IFE. As for conventional GWLP, where each pulse has a Gaussian temporal profile, $t_{decay}$ is in the range of $20 < t_{decay} < 3000$ fs, [22] and in our simulation, we select $t_{decay} = 200\ fs$.

TABLE I. Parameters selected in our model

| Parameters | Value | Units |
|---|---|---|
| $C_p$ | $8.5\times10^6$ | $J\cdot m^{-3}\cdot K^{-1}$ |
| $\gamma$ | $3\times10^3$ | $J\cdot m^{-3}\cdot K^{-2}$ |
| $g_{ep}$ | $1.5\times10^{18}$ | $J\cdot s^{-1}\cdot m^{-3}\cdot K^{-1}$ |
| $F$ | 4 | $mJ/cm^2$ |
| $t_0$ | $35\times10^{-15}$ | s |
| $R$ | $9.4\times10^{12}$ | $s^{-1}$ |
| $T_c$ | 450 | K |
| $t_1$ | $150\times10^{-12}$ | s |
| $\kappa$ | 100 | $J\cdot s^{-1}\cdot m^{-3}\cdot K^{-1}$ |
| $t_{decay}$ | $200\times10^{-15}$ | s |
| $I_0$ | $1.612\times10^{20}$ | $m^{-1}\cdot s^{-1}$ |
| $\beta$ | $2\times10^6$ | $m/A$ |

Employing the model as introduced in the foregoing, the dynamic response of magnetization of the same one sample after excitation with three different polarization SWLPs is surveyed, and the parameters chose for our simulation are listed in table I. During the simulation, the initial magnetization $M = -1$, and this can be gotten easily by an external magnetic field. Figure 2 (a) shows the magnetization dynamics of the same one sample after excitation by three different polarized SWLPs with equal fluence and pulse width. In Fig. 2 (a), we can find the magnetization reversal appears directly when a right-hand circularly polarized SWLP induced on the sample. While the case of $\sigma = 0$ or $\sigma = -1$, no magnetization reversal is observed, and only a rapid demagnetization and slower re-magnetization process exist. This because that the right-hand circularly polarized SWLP will introduce a positive effective magnetic field induced by IFE, which is opposite to the original direction of magnetization. While for the case $\sigma = -1$ will produce a negative field, which is agreement with the original direction of magnetization, and for the case $\sigma = 0$, none magnetic field is introduced. As a result, no reversal can be observed, and only a demagnetization and slower re-magnetization process occur due to the thermal effect. Since the laser energy is a crucial parameter in determining whether the reversal of magnetization will proceed, the dynamics of the magnetization under different laser fluences and the same pulse duration are surveyed. As shown in Fig. 2 (b), when the laser fluence $F < 0.67$ mJ/cm$^2$, no switching is performed, and only a demagnetization and slower re-magnetization process exist. However, with the increase of laser energy, AOS is observed. And further increasing the laser energy, thermal demagnetization appeared as the magnetization is heated up to a high temperature, where the ferromagnet cannot cool down, and the resulting value of $M$ will be 0. In Fig. 2(b), we can get the threshold of AOS for the ferromagnet is about 0.67 mJ/cm$^2$.

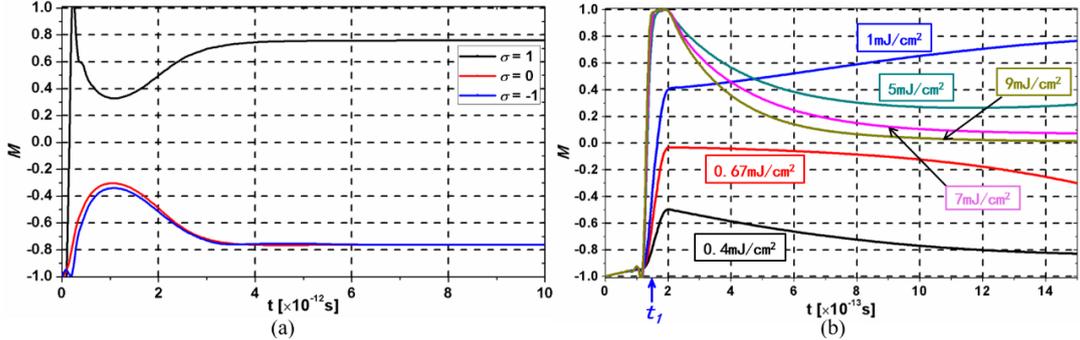

Fig. 2 (a) The magnetization dynamics after excitation by a SWLP with polarization of right-hand circularly polarized ($\sigma = 1$), linearly polarized ($\sigma = 0$), and left-hand circularly polarized ($\sigma = -1$), respectively. The laser fluence and pulses width are 4mJ/cm$^2$ and 35 *fs*, respectively. (b) The magnetization dynamics after excitation by a right-hand circularly polarized laser pulse with 6 different laser fluences, and the laser fluences are 0.4 mJ/cm$^2$, 0.67 mJ/cm$^2$, 1 mJ/cm$^2$, 5 mJ/cm$^2$, 7 mJ/cm$^2$, and 9 mJ/cm$^2$, respectively. Note: the 6 laser pulses have the same pulse width, which is 35 *fs*.

Comparing with the traditional GWLP, we also confirm the superiority of using a SWLP to excite the reversal of magnetization. As far as we know, a GWLP has the temporal profile of $P_{GWLP}(t) = I_0 \cdot F \cdot exp\left(-\left[(t-t_2)/t_0\right]^2\right)$, where $F$, $t_0$, and $t_2$ are the total fluence, FWHM and the center of the pulse duration, respectively. [19] We also define another two parameters $F_{eff}(t) = P(t)/I_0$ and $\Delta\tau$ to study the evolution of laser pulse energy and magnetization reversal. As shown in Fig. 3, $\Delta\tau$ is the time from the center of the SWLP or GWLP to reach the maximum demagnetization of the ferromagnet. Figure 3 shows the magnetization dynamics after excitation by three right-hand circularly polarized laser pulses with the same laser energy $W = 4\times10^8 J \cdot s/m^2$, where $W(t) = \int_0^t P(t)dt$. In Fig. 3 (a) and Fig. 3 (b), the laser pulses are square waves, and the pulses width are 35 $fs$ and $\sqrt{\pi}\times 35\,fs=62.04\,fs$, respectively. $F_{eff}$ is a const, and they are $\sqrt{\pi}\times 40\,J/m^2=70.9\,J/m^2$ and 40 $J/m^2$, respectively. Fig. 3 (c) shows the magnetization dynamics after excitation by a right-hand circularly polarized GWLP, and the pulse width and the peak value of $F_{eff}$ are 35 $fs$ and 40 $J/m^2$, respectively. In Fig. 3 (a), (b) and (c), $\Delta\tau$ is equal to $3.03\times10^{-14}$ $s$, $5.05\times10^{-14}$ $s$ and $9.09\times10^{-14}$ $s$, respectively. Compare with Fig 3 (a) and Fig. 3 (c), we can see that a SWLP is more suitable for the exciting of AOS, and the switching time of AOS is three times faster than the using of a GWLP under the same laser energy and pulse duration. In Fig. 3 (a) and Fig. 3 (b), we can also find that with the increase of the pulse duration, the magnetization dynamics will slow down.

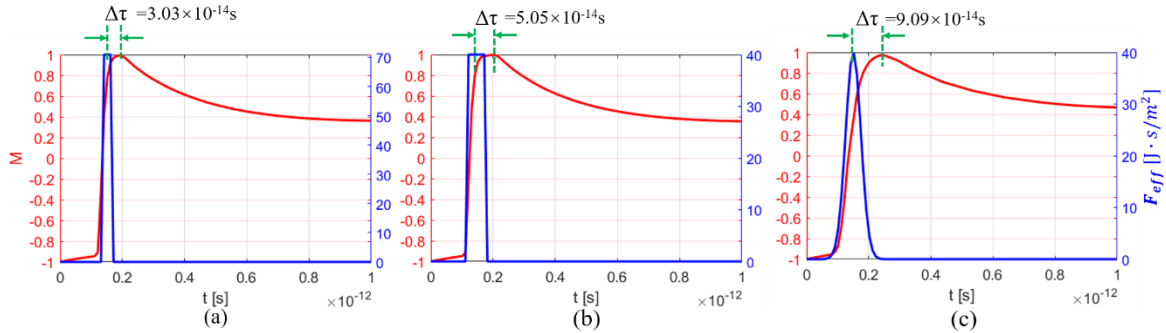

Fig. 3 The magnetization dynamics after excitation by three different laser pulses with right-hand circularly polarized. (a) A SWLP with a pulse width of $t_0$ = 35 $fs$, and the maximum value of $F_{eff}$ is about 70 $J/m^2$; (b) A SWLP with a pulse width of $t_0$ = 62.04 $fs$, and the maximum value of $F_{eff}$ is 40 $J/m^2$; (c) A GWLP with a pulse width of $t_0$ = 35 $fs$, and the maximum value of $F_{eff}$ is 40 $J/m^2$.

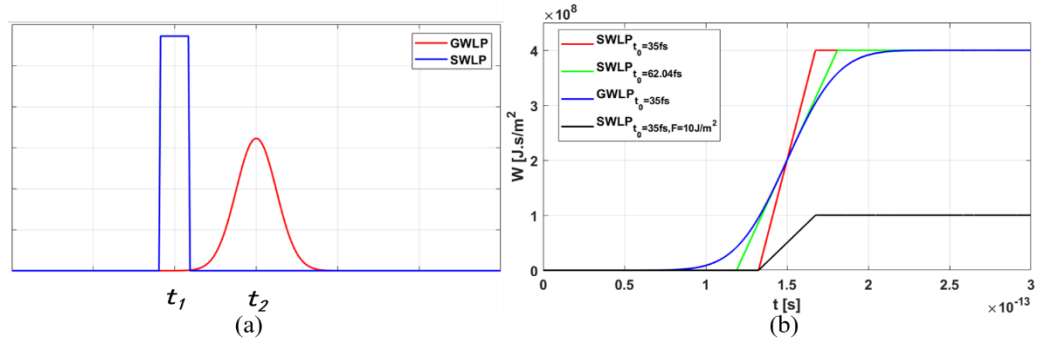

Fig. 4 (a) Temporal profile of a SWLP and a GWLP, and they have the same pulse width of $t_0$ = 35 $fs$. $t_1$ and $t_2$ are the center of the two pulses duration; (b) Evolution of laser energy $W$(t) of four different laser pulses. The first three laser pulses, which are a SWLP, a SWLP, and a GWLP, have the same laser fluence, and their pulse width are 35 $fs$, 62.04 $fs$ and 35 $fs$, respectively. The last pulse is a SWLP, and the laser fluence is 10 $J/m^2$ with pulse width 35$fs$.

We also explore the mechanism that why the reversal of AOS speed up, if a SWLP is selected as the heat source. Figure 4 (a) shows the temporal profile of a SWLP and a GWLP, where they have the same pulse width and laser energy, and we can see the rising or falling edge of the SWLP is sharp, and we believe that the accumulate of energy is rapider than the GWLP, which have a gently rising and falling. Figure 4 (b) confirms our assumption. In Fig. 4 (b), the laser energy of the first three laser pulses, which are a SWLP, a SWLP, and a GWLP, is $4\times10^8 J\cdot s/m^2$, and their pulse width are 35 $fs$, 62.04 $fs$ and 35 $fs$, respectively. The last pulse is a SWLP with laser energy of $1\times10^8 J\cdot s/m^2$, which is corresponding to $F$ = 10 $J/m^2$, and the pulse is 35 $fs$. As shown in Fig. 4 (b), the rate of increase of laser energy, which is corresponding to the slope of $W$(t), is different, and under the same laser fluence, a SWLP has a larger slope than a GWLP. We can also find that, a SWLP with a shorter pulse-width will have a larger "heat accumulating effect" due to the larger peak value. From the analysis in the foregoing, we can confirm that "heat accumulating effect" play an important role during AOS.

In conclusion, we survey the AOS in a ferromagnet of Co/Pt system by a SWLP vie the combination of heating and IFE. Simulation result shows that the switching time of magnetization is 3 times faster than the using of a traditional GWLP under the same laser energy and pulse duration, and to further speed up the magnetization reversal, using a SWLP can be an effective candidate. We also predict that AOS for the ferromagnet in our model is possible with laser energy larger than 0.67 mJ/cm$^2$. At last, we furthermore demonstrate that the "heat accumulating effect" of a laser-pulse is an important factor, which influences the AOS, and a SWLP has a larger effect of heat accumulating than a GWLP.